\documentclass[11pt,Times,aps,shownopacs,superscriptaddress,nofootinbib,preprint]{revtex4}
\usepackage{graphics}
\usepackage{graphicx}
\usepackage{epsfig}
\usepackage{psfrag}
\usepackage{color}
\usepackage{axodraw}

\usepackage{amsfonts}
\usepackage{amssymb}

\newcommand{\be}{\begin{equation}}
\newcommand{\ee}{\end{equation}}
\newcommand{\bea}{\begin{eqnarray}}
\newcommand{\eea}{\end{eqnarray}}
\newcommand{\ba}{\begin{array}}
\newcommand{\ea}{\end{array}}
\newcommand{\bmat}{\left(\ba}
\newcommand{\emat}{\ea\right)}

\newcommand{\ler}{\stackrel{\scriptstyle <}{\scriptstyle\sim}}
 \newcommand{\ger}{\stackrel{\scriptstyle >}{\scriptstyle\sim}}

\newcommand*{\e}{\epsilon}
\providecommand*{\ler}{\stackrel{\scriptstyle <}{\scriptstyle \sim}}
\providecommand*{\ger}{\stackrel{\scriptstyle >}{\scriptstyle \sim}}
\renewcommand{\le}{\left(}
\newcommand{\ri}{\right)}
\newcommand{\cov}{{\cal{D}}}

\def\lm{{\cal L}}
\def\row{\rightarrow}

\def\Si{\Psi}

\newcommand{\LL}{{\mbox{\scriptsize LL}}}
\newcommand{\LR}{{\mbox{\scriptsize LR}}}

\newcommand{\RR}{{\mbox{\scriptsize RR}}}

\newcommand{\elm}{{\mbox{\scriptsize em}}}

\newcommand{\unity}{{\hbox{1\kern-.8mm l}}}

\def\321{$SU(3)\times SU(2)\times U(1)$}

\def\a{\alpha}
\def\b{\beta}
\def\g{\gamma}

\def\e{\epsilon}
\def\l{\lambda}
\def\n{\nu}
\def\m{\mu}

\def\th{\theta}

\def\dm21{\Delta {\mbox m}^2_{21}}
\def\dm32{\Delta {\mbox m}^2_{32}}
\def\dm{\Delta {\mbox m}^2}

\newcommand{\comment}[1]{}

\newcommand{\cref}[1]{Chapter~\ref{c.#1}}

\def\pa{\partial}

\def\lsim{\raise0.3ex\hbox{$\;<$\kern-0.75em\raise-1.1ex\hbox{$\sim\;$}}} 
\def\gsim{\raise0.3ex\hbox{$\;>$\kern-0.75em\raise-1.1ex\hbox{$\sim\;$}}}

\begin{document}

\title{Introduction to the MSSM} 
\author{Sudhir K. Vempati}
\affiliation{Centre for High Energy Physics, Indian Institute of
Science, Bangalore 560 012, India}
\email{vempati@cts.iisc.ernet.in}
\begin{abstract}
These lecture notes are based on  a first course on the Minimal Supersymmetric
Standard Model. The level of the notes is introductory and pedagogical.  Standard
Model, basic supersymmetry algebra and its representations are considered
as prerequisites. The topics covered include particle content, structure of the 
lagrangian, supersymmetry breaking soft terms, electroweak
symmetry breaking and the sparticle mass spectrum.  Popular supersymmetry
breaking models like minimal supergravity and gauge mediation are also introduced. 
\end{abstract} 
\maketitle

\section{Prerequisites}
These lectures\footnote{ Based on lectures presented at SERC school,  held at IIT-Bombay,  Mumbai. }
are devised as an introduction to the Minimal Supersymmetric
Standard Model. In the course of these lectures, we will introduce the
basic features of the Supersymmetric Standard Model, the particle content,
the structure of the lagrangian, feynman rules, supersymmetric breaking
soft terms, Electroweak  symmetry breaking and the mass spectrum of the
MSSM. Supersymmetry is a vast subject and these lectures are definitely
not a comprehensive review and in fact, they are also not what one could
term as an introduction to supersymmetric algebra and supersymmetric gauge
theories. The prerequisites for this course are  a good knowledge
of the Standard Model and also supersymmetry say, at the level of first eight
chapters of Wess and Bagger\cite{wessbagger}: 
supersymmetry transformations, representations, 
superfields, supersymmetric gauge theories and elements of supersymmetry 
breaking. It is strongly recommended that readers keep a text book on 
basic course of supersymmetry \cite{westetal,sohnius,onethousand} with them all the time for 
consulting, while going  through this lecture notes. 

The lectures are organised as follows : in the next section, we will give
a lightening introduction to the Standard Model and the structure of its
lagrangian. The reason for this being that we will like to introduce the
MSSM (Minimal Supersymmetric Standard Model ) in a similar organisational
fashion, which makes it easier to remember the MSSM lagrangian - as well
as arranging the differences and similarities , one expects in the 
supersymmetric theories in a simpler way, if possible. The next section
would introduce the basic form of the MSSM lagrangian - the three functions
of the chiral/vector superfields - the superpotential, the Kahler potential 
and the field
strength superfield and the particle spectrum. The fourth section will be
devoted to R-parity and some sample feynman rules. Supersymmetry breaking
and electroweak symmetry breaking will be introduced in section 5 and 
the physical supersymmetric particle mass spectrum will be done in section 6. 
Higgs sector will be reviewed in section 7, while we close with some `standard'
models of supersymmetry breaking in section 8.

Finally for the students not completely familiar with Standard Model,
we point out at some references with increasing order of difficulty
in reading and requirements of pre-requisites. These are : 
(a) Aitchison and Hey\cite{athey}, Gauge Field Theories, Vol I and Vol II 
(b)A good functional introduction to field theory required for 
understanding Standard Model can be found in :  M. Srednicki, 
Quantum Field Theory \cite{srednicki} (c) M. E. Peskin and
D. Schroeder, Quantum Field Theory \cite{peskin} (d) E. Abers and B. W. Lee,
Physics Reports on Gauge Field Theories \cite{aberslee} (e) T. Cheng and
L. Li, Gauge theory of elementary particle physics\cite{chengli} 
(f) S. Weinberg, Quantum  Theory of Fields, Vol I -II\cite{weinberg}, 
(g) S. Pokorski, Gauge Field Theories \cite{pokorski} and 
(h) Donoghue, Golowich and Holstein, Dynamics of the 
Standard Model \cite{dynamics}.

\section{Step 0 :  A lightening recap of the Standard Model}
The Standard Model (SM) is a spontaneously broken Yang-Mills 
quantum field theory describing the strong and electroweak 
interactions.  The theoretical assumption on
which the Standard Model rests on is the principle of local gauge
invariance with the gauge group given by
\begin{equation} 
\label{gsm}
G_{SM} \equiv SU(3)_c \times SU(2)_L \times U(1)_Y,
\end{equation}
where the subscript $c$ stands for color,  $L$ stands for the
`left-handed' chiral group whereas $Y$ is the hypercharge. 
The particle spectrum and their transformation properties
under these gauge groups are given as,
\bea
\label{smspectrum}
Q_i \equiv \bmat{c} {u_L}_i \\ {d_L}_i \emat \sim
\le 3,~ 2,~ {1 \over 6} \ri
&\;\;\;\;& U_i \equiv {u_R}_i \sim \le \bar{3},~ 1,~ {2 \over 3} \ri \nonumber \\
&\;\;\;\;& D_i \equiv {d_R}_i \sim \le \bar{3},~ 1,~ -{1 \over 3}\ri \nonumber \\
L_i \equiv \bmat{c} {\n_L}_i \\ {e_L}_i \emat \sim \le 1,~ 2,~ -{1 \over 2} \ri
&\;\;\;\;& E_i \equiv {e_R}_i \sim  \le 1,~ 1,~ -1 \ri \nonumber 
\eea
In the above $i$ stands for the generation index, which runs over the
there generations $i=1,2,3$.  $Q_i$ represents the 
left handed quark doublets containing both the up and down quarks of 
each generation. Similarly, $L_i$ represents left
handed lepton doublet, $U_i,~D_i,~E_i$ represent right handed up-quark,
down-quark and charged lepton singlets respectively.
The numbers in the parenthesis represent the transformation properties 
of the particles under $G_{SM}$ in the order given in eq.(\ref{gsm}).
For example, the quark doublet $Q$ transforms a triplet (3) under 
$SU(3)$ of strong interactions, a doublet (2) under weak interactions 
gauge group and carry a hypercharge $(Y/2)$ of 1/6 \footnote{Note that 
the hypercharges
are fixed by the Gellman-Nishijima relation $Y/2~ =~ Q - T_3 $, where $Q$
stands for the charge of the particle and $T_3$ is the eigenvalue of the 
third generation of the particle under $SU(2)$.}.  
In addition to the fermion spectra represented above, there is also
a fundamental scalar called Higgs whose transformation properties
are given as
\be
\label{smhiggs}
H \equiv \bmat{c} H^+ \\ H^0 \emat \sim \le 1,~2,~1/2 \ri.
\ee

However, the requirement of local gauge invariance will not be fulfilled unless
one includes the gauge boson fields also. Including them, the 
total lagrangian with the above particle spectrum and gauge group can
 be represented as,

\be
\lm_{SM} = \lm_F + \lm_{YM}  + \lm_{yuk} + \lm_S.
\ee

\noindent
The fermion part $\lm_F$ gives the kinetic terms for the fermions
as well as their interactions with the gauge bosons. It is given as,
\be
\label{lfer}
\lm_{F} = i \bar{\Si} \g^\m \cov_\m \Si, 
\ee
where $\Si$ represents all the fermions in the model, 
\be
\Si = \le Q_i~U_i,~ D_i,~ L_i,~ E_i \ri
\ee
where $\cov_\m$ represents the covariant derivative of the field given as,
\be
\label{cova}
\cov_\m  = \partial\m - i g_s G_\m^A \l^A - i{g \over 2} W_\m^I \tau^I
- i g' B_\m Y
\ee
Here $A= 1,..,8$ with $G_\m^A$ representing the $SU(3)_c$ gauge bosons,
$I = 1,2,3$ with $W_\m^I$ representing the $SU(2)_L$ gauge bosons. The
$U(1)_Y$ gauge field is represented by $B_\m$. The kinetic terms for 
the gauge fields and their  self interactions  are given by,
\be
\label{lym}
\lm_{YM} = - { 1\over 4} G^{\m\n A} G_{\m\n}^A -{1 \over 4} W^{\m\n I}
W_{\m\n}^I - {1 \over 4} B^{\m\n} B_{\m\n}
\ee
with
\bea
G_{\m\n}^A&=&\pa_\m G_\n^A - \pa_\n G_\m^A + g_s ~f_{ABC}
G_\m^B G_\n^C \nonumber \\
F_{\m\n}^I&=&\pa_\mu W_\n^I - \pa_\n W_\m^I + g~ f_{IJK} W_\m^J W_\n^K \nonumber\\
B_{\m\n}&=& \pa_\m B_\n - \pa_\n B_\m ,
\eea
where $f_{ABC (IJK)}$ represent the structure constants of the
$SU(3)(SU(2))$ group.

In addition to the gauge bosons, the fermions also interact 
with the Higgs boson, through the dimensionless Yukawa couplings
given by
\be
\label{lyuk}
\lm_{yuk} = h^u_{ij} \bar{Q}_i U_j \tilde{H} + h^d_{ij} \bar{Q}_i D_j H +
h^e_{ij} \bar{L}_i E_j H + H.c
\ee
where $\tilde{H} = i \sigma^2 H^\star$. 
These couplings are responsible for the fermions to attain masses once
the gauge symmetry is broken from 
$G_{SM}~ \rightarrow~ SU(3)_c \times U(1)_{em}$. 
This itselves is achieved by the scalar part of the lagrangian 
which undergoes spontaneous symmetry breakdown. 
The scalar part of the lagrangian is given by,
\be
\label{lscalar}
\lm_{S} =   \le \cov_\m H \ri^{\dagger} \cov_\m H - V(H),
\ee where
\be
V(H)  =  \m^2 H^\dagger H + \l  \le H^\dagger H \ri^2
\ee
For $\mu^2~ <~0$, the Higgs field attains a vacuum expectation value 
({\it vev}) at the minimum of the potential. The resulting goldstone 
bosons are `eaten
away' by the gauge bosons making them massive through the so-called 
Higgs mechanism. Only one degree of the Higgs field remains physical,
the only scalar particle of the SM - the Higgs boson. The fermions
also attain their masses through their Yukawa couplings, once the 
Higgs field attains a {\it vev}. The only exception is the neutrinos
which do not attain any mass due to the absence of right handed neutrinos
in the particle spectrum and thus the corresponding Yukawa couplings. 
Finally, the Standard Model is renormalisable and anomaly free. We 
would also insist that the Supersymmetric version of the Standard Model
keeps these features of the Standard Model intact.

\subsubsection{Think it Over}
\noindent
Here are some important aspects of the Standard Model which have not
found a mention in the above. These are formulated in some sort of
a problem mode, which would require further study. 
\begin{itemize}
\item What is the experiment that showed that there are only
three generations of particles in the Standard Model ? Can one
envisage a fourth generation ? If so, what are the constraints
this generation of particles expected to satisfy ? 
\item The gauge bosons `mix' at the tree level by an angle $\tan~\theta_W =
({g'\over g})^2$. What happens at the 1-loop level ? However are the
relevant observables classified ? Some relevant information can be 
found at\cite{hollik}. 
\item What are the theoretical limits on the Higgs boson mass ?  
How sensitive is the upper bound on the Higgs mass from precision 
measurements to the top quark mass ? What is the lower limit on 
the Higgs mass from the LEP experiment ? Some relevant information
can be found at\cite{lephwg}.  
\item The LHC experiment has been rapidly constraining the 
allowed parameter space of the Higgs boson. For latest information
have a look at \cite{cerntwiki}.  
\item 
What is the CKM mixing ? How well are these angles measured ? 
What is the present status after the results from various B-factories
about the CP phase in the SM ? What is the analogous mixing in the
leptonic sector called ? In comparison to the CKM matrix, how well
are these angles measured ? 
\end{itemize}


\section{Step 1 :  Particle Spectrum of MSSM }

What we aim to build over the course of next few lectures is a supersymmetric
version of the Standard Model, which means the lagrangian we construct should
not only be gauge invariant under the Standard Model gauge group $G_{SM}$ but 
also now be supersymmetric invariant. Such a model is called Minimal 
Supersymmetric Standard Model with the the word 'Minimal' referring to 
minimal choice of the particle spectrum required to make it work. Furthermore,
we would also like the MSSM to be renormalisable and anomaly free, just like
the Standard Model is. 

Before we proceed to discuss about the particle spectrum, let us remind 
ourselves that ordinary quantum fields are upgraded in supersymmetric\footnote{All
through this set of lectures, whenever we mention supersymmetry we mean N=1
SUSY ; only one set of SUSY generators.} theories to so-called supermultiplets or 
superfields\footnote{Superfields are functions (fields) written over 
a `superspace' made of ordinary space ($x_\mu$) and two fermionic `directions' 
($\theta$,$\bar{\theta}$); they are made up of quantum fields whose spins
differ by 1/2.  To build interaction lagrangians one normally 
resorts to this formalism, originally given by Salam and Strathdee\cite{salamstrahdee}, as 
superfields simplify addition and multiplication of the representations.
It should be noted however that the component fields 
may always be recovered from superfields by a power series expansion
in grassman variable, $\theta$. \\
A chiral superfield has an expansion :
\be
\Phi = \phi + \sqrt{2} \theta \psi + \theta \theta F,
\ee 
where $\phi$ is the scalar component, $\psi$, the two component spin 1/2 fermion
and $F$ the auxiliary field. A vector superfield in (Wess-Zumino gauge) has
an expansion :
\be
V = -\theta \sigma^\mu \bar{\theta} A_\mu + i \theta \theta \bar{\theta} \bar{\lambda} - i \bar{\theta} \bar{\theta} \theta \lambda + {1 \over 2} \theta \theta \bar{\theta} \bar{\theta} D 
\ee }.
Given that supersymmetry transforms a fermion into a boson and vice-versa, 
supermultiplets or superfields are multiplets which collect fermion-boson
pairs which transform in to each other.  We will deal with two kinds of 
superfields - vector superfields and chiral superfields.  A chiral 
superfield\footnote{Here we are presenting the particle content in the 
off-shell formalism.} contains a weyl fermion,  a scalar and and an 
auxiliary scalar field generally denoted by F. A vector superfield contains 
a spin 1 boson, a spin 1/2 fermion and an auxiliary scalar field called D. 

The minimal supersymmetric extension of the Standard Model 
is built by replacing every standard model matter field
by a chiral superfield and every vector field by a vector superfield.
Thus the existing particle spectrum of the Standard Model is doubled. 
The particle spectrum of the MSSM and their 
transformation properties under $G_{SM}$ is given by,
\bea
\label{mssmspectrum}
Q_i \equiv \bmat{cc} u_{L_i} & {\tilde u}_{L_i} \\ 
d_{L_i} & \tilde{d}_{L_i} \emat \sim 
\le 3,~2,~{1 \over 6} \ri
&\;\;& U_i^c \equiv \bmat{cc} u_i^c  & \tilde{u}^{c}_i \emat
 \sim \le \bar{3},~ 1,~ -{2 \over 3} \ri \nonumber \\
&\;\;& D_i \equiv \bmat{cc} d_i^c & \tilde{d}^{c}_i  \emat
\sim \le \bar{3},~ 1,~ {1 \over 3} \ri \nonumber \\
L_i \equiv \bmat{cc} \n_{L_i} & \tilde{\n}_{L_i}  \\ e_{L_i}& 
\tilde{e}_{L_i} \emat 
\sim \le 1,~ 2,~ -{1 \over 2} \ri
&\;\;& E_i \equiv \bmat{cc} e_i^c & \tilde{e}^{c}_i \emat \sim 
 \le 1,~ 1,~ 1 \ri \nonumber \\
\eea
The scalar partners of the quarks and the leptons are typically named
as `s'quarks and `s'leptons. Together they are called sfermions. 
For example, the scalar partner of the top
quark is known as the `stop'. In the above, these are represented by
a `tilde' on their SM counterparts. As in the earlier case, 
the index $i$ stands for the generation index. 

There are two distinct features in the spectrum of MSSM : (a) Note
that we have used the conjugates of the right handed particles, instead
of the right handed particles themselves. There is no additional conjugation
on the superfield itselves, the $c$ in the superscript just to remind ourselves
that this chiral superfield is made up of conjugates of SM quantum fields.
In eq.(\ref{mssmspectrum}), $u^c~=~u_R^\dagger$ and 
$\tilde{u}^c~=~\tilde{u}_R^\star$.
This way of writing down the particle spectrum is highly useful for
reasons to be mentioned later in this section. Secondly (b) At least
two Higgs superfields are required to complete the spectrum - one giving
masses to the up-type quarks and the other giving masses to the down type
quarks and charged leptons. As mentioned earlier, this is the minimal number
of Higgs particles required for the model to be consistent from a quantum
field theory point of view\footnote{The Higgs field has a fermionic partner,
higgsino which contributes to the anomalies of the SM. At least two such 
fields with opposite hyper-charges ($U(1)_Y$) should exist to cancel the
anomalies of the Standard Model.}. These two Higgs superfields have the following
transformation properties under $G_{SM}$: 
\bea
\label{higgssp}
H_1 &\equiv& \bmat{cc} H^{0}_1 & \tilde{H}_1^0 \\ H^{-}_1 & \tilde{H}_1^{-} 
\emat 
\sim \le 1,~2,~-{1 \over 2} \ri  \nonumber \\
H_2 &\equiv & \bmat{cc} H^+_2 & \tilde{H}_2^+  \\ H^0_2 & \tilde{H}_2^0 \emat
\sim \le 1,~2,~{1 \over 2} \ri 
\eea
The Higgsinos are represented by a~$\tilde{}$~on them. This completes the
matter spectrum of the MSSM. Then there are the gauge bosons and their 
super particles. Remember that in supersymmetric theories, the gauge symmetry
is imposed by the transformations on matter superfields as  :
\be
\Phi' = e^{i \Lambda_l t_l } \Phi 
\ee 
where $\Lambda_l$ is an arbitrary chiral superfield and $t_l$ represent the
generators of the gauge group which are $l$ in number and the index $l$ is
summed over\footnote{To  be more specific, $t_l$ is just a number for the
abelian groups. For non-abelian groups, $t_l$ is a matrix and so is $\Lambda_l$, 
with $\Lambda_{ij} = t^l_{ij} \Lambda_l$ Note that $V$ is also becomes a matrix
in this case.}. The gauge invariance
is restored in the kinetic part by introducing a (real) vector superfield, $V$ 
such that the combination 
\be
\Phi^\dagger e^{ gV} \Phi
\ee 
remains gauge invariant. For this to happen, the vector superfield $V$ itselves
transforms under the gauge symmetry as 
\be
\delta V = i(\Lambda - \Lambda^\dagger)
\ee 
The supersymmetric invariant kinetic part 
of the lagrangian is given by:
\be
\label{kahlerp}
\mathcal{L}_{kin} = \int d\theta^2 d\bar{\theta}^2 \Phi^\dagger e^{gV} \Phi = 
 \Phi^\dagger e^{gV} \Phi |_{\theta\theta\bar{\theta}\bar{\theta}}
\ee
In the MSSM, corresponding to three  gauge groups of the SM and for
each of their corresponding gauge bosons, we need to add a vector superfield
which transforms as the adjoint under the gauge group action.  Each vector superfield
contains the gauge boson and its corresponding super partner called gaugino.
Thus in MSSM we have the following vector superfields and their corresponding
transformation properties under the gauge group, completing the particle
spectrum of the  MSSM:
\bea
V_s^A & : & \bmat{cc} G^{\m A} & \tilde{G}^A \emat ~~ \sim~~ (8,1,0) \nonumber \\
V_W^I & : & \bmat{cc} W^{\m I}& \tilde{W}^I \emat ~~~\sim~~ (1,3,0)  \nonumber \\
\label{vsf}
V_Y & :  &\bmat{cc} B^{\m} & \tilde{B} \emat~~\sim~~(1,1,0)
\eea
The $G$'s ($G$ and $\tilde{G}$) represent the gluonic fields and their 
superpartners called gluinos,  the index $A$ runs from $1$ to $8$. The $W$'s 
are the $SU(2)$ gauge bosons and their superpartners `Winos', the index $I$ 
taking values from $1$ to $3$ and finally $B$s represents the $U(1)$ gauge
boson and its superpartner `Bino'. Together all the superpartners of the
gauge bosons are called `gauginos'. This completes the particle spectrum 
of the MSSM.

\section{Step 2: The superpotential and R-parity}

The supersymmetric invariant lagrangian is constructed from functions of 
superfields. In general there are three functions which are: 
(a) The K{\"a}hler potential, $K$, which is a real
function of the superfields (b) The superpotential $W$, which is a
holomorphic (analytic) function of the superfields, and (c) the gauge
kinetic function $f_{\alpha \beta}$ which appears in supersymmetric
gauge theories. This is the coefficient of the product of field strength
superfields,  $\mathcal{W}_\alpha \mathcal{W}^\beta$. 
The field strength superfield is derived from the vector superfields
contained in the model. $f_{\alpha \beta}$ determines
the normalisation for the gauge kinetic terms.  
In MSSM,  $f_{\alpha \beta} = \delta_{\alpha \beta}$. 
The lagrangian of the MSSM is thus given in terms of 
 $G_{SM}$ gauge invariant functions
$K$, $W$ and add the field strength superfield  $\mathcal{W}$, 
for each of the vector superfields in the spectrum. 

The gauge invariant K{\"a}hler potential has already been discussed
in the eqs.(\ref{kahlerp}). For the MSSM case, the K{\"a}hler potential
will contain all the three vector superfields corresponding to the $G_{SM}$
given in the eq.(\ref{vsf}). Thus we have :
\be
\label{mssmkahler}
\mathcal{L}_{kin} = \int d\theta^2 d\bar{\theta}^2 \sum_{\scriptstyle{\text{SU(3)},\text{SU(2)},\text{U(1)}}} 
\Phi_\beta^\dagger~ e^{ gV} \Phi_\beta  
\ee 
where the index $\beta$ runs over all the matter fields 
$\Phi_\beta~ =~ \{Q_i, U^c_i, D^c_i, L_i, e^c_i, H_1, H_2 \}$\footnote{The 
indices $i,j,k$ always stand for the three generations through out this
notes, taking values between 1 and 3.} in appropriate representations.
Corresponding to each of the gauge groups in 
$G_{SM}$, all the matter fields which transform non-trivially under this
gauge group\footnote{As given in the list of  representations in 
eqs. (\ref{mssmspectrum},\ref{higgssp}) } are individually taken and 
the grassman ($d\theta^2 d{\bar{\theta}}^2$) integral is evaluated with 
the corresponding vector superfields in the exponential\footnote{Remember
that the function $e^{gV}$ truncates at ${1 \over 2} g^2 V^2$ in the 
Wess-Zumino gauge. In fact, in this gauge, this function can be determined
by noting:
\be
\exp V_{WZ} = 1 - \theta \sigma^\mu \bar{\theta} A_\mu + i \theta \theta 
\bar{\theta} \bar{\lambda} - i \bar{\theta} \bar{\theta} \theta \lambda +
{1 \over 2}~ \theta \theta \bar{\theta} \bar{\theta}~ ( D - {1 \over 2}
 A^\mu A_\mu),
\ee
for an abelian Vector superfield. Here as usual $\lambda$ denotes the
gaugino field while $A_\mu$ represents the gauge field. D represents
the auxiliary field of the Vector multiplet. The extension to the 
non-abelian case is straight forward.}. After expanding and evaluating 
the integral, we get the 
lagrangian which is supersymmetric invariant in terms of the ordinary 
quantum fields - the SM particles and the superparticles. This part of 
the lagrangian would give us the kinetic
terms for the SM fermions, kinetic terms for the sfermions and their 
interactions with the gauge bosons and in addition also the interactions 
of the type: fermion-sfermion-gaugino which are structurally like the  
Yukawa interactions ($ff\phi$), but carry gauge couplings. 
Similarly, for the Higgs fields, this part of the lagrangian gives
the kinetic terms for the Higgs fields and their fermionic superpartners
Higgsinos and the interaction of the gauge bosons with the Higgs fields
and Higgs-Higgsino-gaugino vertices. 

The second possible function of the superfields is the analytic 
or holomorphic function\footnote{This would mean that $W$ is purely
a function of complex fields ($z_1z_2z_3$) or its conjugates ($z_1^\star z_2^\star z_3^\star$).  }
of the superfields called the superpotential, $W$. This function essentially
gives the interaction part of the lagrangian which is independent of the
gauge couplings, like the Yukawa couplings.  If renormalisability 
is demanded, the dimension of the superpotential is restricted to be
less than or equal to three, $[W]~ \leq~ 3 $~ \textit{i.e,}~ only products of 
three or less number of chiral superfields are allowed. Imposing this 
restriction of renormalisability the most general  $G_{SM}$ gauge 
invariant form of the $W$ for the matter spectrum of MSSM (\ref{mssmspectrum},\ref{higgssp})  is given as 
\be
\label{super}
W  =  W_1 + W_2, \\
\ee
\noindent
where
\bea
\label{superyuk}
W_1 &=& h^u_{ij} Q_i U_j^c H_2 + h^d_{ij} Q_i D_j^c H_1 + 
h_{ij}^e L_i E_j^c H_1 + \m H_1 H_2 \\
\label{superrpv}
W_2&=& \e_i L_i H_2 + \l_{ijk} L_i L_j E_k^c + \l'_{ijk} L_i Q_j D_k^c 
+ \l''_{ijk} U_i^c D_j^c D_k^c .
\eea
Here we have arranged the entire superpotential in to two parts, $W_1$ and
$W_2$ with a purpose. Though both these parts are gauge invariant, $W_2$ 
also violates the global  lepton number and baryon quantum numbers.
The simultaneous presence of both these set of operators can lead to rapid
proton decay and thus can make the MSSM phenomenologically invalid. For 
these reasons, one typically imposes an additional symmetry called R-parity
in MSSM which removes all the dangerous operators in $W_2$. We will deal 
with R-parity in greater detail in the next section. For the present, let
us just set $W_2$ to be zero due to a symmetry called R-parity and just
call $W_1$ as $W$. The lagrangian can be derived from the superpotential
containing (mostly) gauge invariant product of the three superfields by
taking the $\theta\theta$ component, which can be represented in the integral
form as
\be
\mathcal{L}_{yuk}  = \int d\theta^2~~ W (\Phi) + 
\int d\bar{\theta}^2~~ \bar{W}(\bar{\Phi})
\ee 
This part gives\footnote{The $\theta\theta$ components of the product of three
chiral superfields is given as\cite{wessbagger}
\be
\Phi_i\Phi_j\Phi_k|_{\theta\theta} = -\psi_i\psi_j \phi_k - \psi_j\psi_k \phi_i 
 -\psi_k\psi_i \phi_j + F_i \phi_j \phi_k  + F_j \phi_k \phi_i + F_k \phi_i \phi_j,
\ee 
where as earlier, $\psi_i$ represents the fermionic, $\phi_i$ the scalar 
and $F_i$ the 
auxiliary component of the chiral superfield $\Phi_i$. Similarly for the
product of two superfields on has : 
\be
\Phi_i\Phi_j|_{\theta\theta} = -\psi_i\psi_j  
+  F_i \phi_j + F_j \phi_i
\ee } 
the standard Yukawa couplings for the fermions with the Higgs, in addition
also give the fermion-sfermion-higgsino couplings and scalar terms. 
For example, the coupling $h^u_{ij}~Q_i u_j^c H_2$ in the superpotential
has the following expansion in terms of the component fields : 
\bea 
\mathcal{L}_{yuk} &\supset& h^u_{ij}~Q_i u_j^c H_2 ~|_{\theta\theta} \nonumber\\
 &\supset& h^u_{ij}~ (~u_i u_j^c H_2^0 - d_i u_j^c H_2^+~)
 ~|_{\theta\theta}\nonumber \\
&\supset& h^u_{ij}(\psi_{u_i} \psi_{u_j^c} \phi_{H_2^0} + \phi_{\tilde{u}_i} 
\psi_{u_j^c} \psi_{\tilde{H}_2^0} + 
\psi_{u_i} \phi_{\tilde{u}_j^c} \psi_{\tilde{H}_2^0} - 
\psi_{d_i} \psi_{ u_j^c} \phi_{H_2^+} - 
\phi_{\tilde{d}_i} \psi_{u_j^c} \psi_{\tilde{H}_2^+} - \psi_{d_i} 
\phi_{\tilde{u}_j^c} \psi_{ \tilde{H}_2^+}) \\
 &\equiv& h^u_{ij}~ (~u_i u_j^c H_2^0 + ~\tilde{u}_i u_j^c \tilde{H}_2^0 + 
u_i \tilde{u}_j^c \tilde{H}_2^0 - d_i u_j^c H_2^+~ - 
\tilde{d}_i u_j^c \tilde{H}_2^+~ - d_i \tilde{u}_j^c \tilde{H}_2^+~),
\eea
where in the last equation, we have used the same notation for the chiral 
superfield as well as for its lowest component namely the scalar component.
Note that  we have not written the F-terms which give rise to the scalar terms
in the potential. 
Similarly, there is the $\mu$ term which gives `Majorana' type
mass term for the Higgsino fields. 

Finally, for every vector superfield (or a set of superfields) we 
have an associated field strength superfield $\mathcal{W}^\alpha$,
which gives the kinetic terms for the gauginos and the field strength
tensors for the gauge fields. Given that it is a chiral superfield, 
the component expansion is given by taking the $\theta\theta$
component of `square' of that superfield\footnote{In the Wess-Zumino gauge, 
$\mathcal{W}_\alpha = - {1 \over 4} \bar{\mathcal{D}} \bar{\mathcal{D}}
\mathcal{D}_\alpha V_{WZ}$ 
\cite{wessbagger} ($\mathcal{D}$ is the differential operator on superfields)
and the lagrangian has the form :
\be
\label{gaugekinetic}
\mathcal{L} \supset {1 \over 4} \left( \mathcal{W}^\alpha \mathcal{W}_\alpha 
|_{\theta\theta} +  
\mathcal{W}^{\dot{\alpha}} \mathcal{W}_{\dot{\alpha}}  
|_{\bar{\theta}\bar{\theta}} \right)
= {1 \over 2} D^2 - {1 \over 4} F_{\mu \nu} F^{\mu \nu} - i \lambda 
\sigma^\mu \partial_\mu \bar{\lambda}
\ee
$D$ represents the auxiliary component of the vector superfields. The extension 
to non-abelian vector superfields in straight forward.}. In the MSSM, we have
to add the corresponding field strength $\mathcal{W}$ superfields for 
electroweak vector superfields, $W$ and $B$  
as well as for the gluonic $G$ vector superfields of eqs.(\ref{vsf}). 

So far we have kept the auxiliary fields  ($D$ and $F$) of various chiral 
and vector superfields in the component form of our lagrangian. However, 
given that these fields are unphysical, they have to be removed from the
lagrangian to go ``on-shell". To eliminate the $D$ and $F$ fields, we have
to use the equations of motions of these fields which have simple solutions
for the $F$ and $D$ as :
\be
F_i = { \partial W \over \partial \phi_i} ~~~~ \;;\;~~~~ 
D_A = -g_A ~~ \phi^\star_i~ T^A_{ij} ~\phi_j ,
\ee 
where $\phi_i$ represents all the scalar fields present in MSSM. The index
$A$ runs over all the gauge groups in the model. For example, for $U(1)_Y$,
$T^A_{ij} = ~(Y^2/2) \delta_{ij}$. The $F$ and $D$ terms
together form the scalar potential of the MSSM\footnote{Later we will
see that there are also additional terms which contribute to the scalar
potential which come from the supersymmetry breaking sector.} which is
given as
\be
V =  \sum_i~| F_i~  |^2~ +~ {1 \over 2}~ D^A D_A 
\ee
Putting together, we see that the lagrangian of the MSSM with SUSY unbroken 
is of the form :
\be
\label{mssmlag}
\lm_{MSSM}^{(0)} = \int \left( d \th^2 ~W(\Phi) + H.c \right)  
+ \int d \th^2~ d {\bar \th}^2 ~
\Phi_i^\dagger~ e^{g V}~ \Phi_i + \int 
\left(d \th^2~ {\cal W}^\a {\cal W}_\a + H.c \right).
\ee 
where all the functions appearing in (\ref{mssmlag})  have been discussed in 
eqs.(\ref{mssmkahler},\ref{superyuk}) and (\ref{gaugekinetic}). 

\subsubsection{\textit{Think it over }}
\begin{itemize}

\item The full supersymmetric lagrangian of the Standard Model can be 
constructed from the prescription given in the above section. Identify
the dominant one-loop contributions for the Higgs particle. Note that
SUSY is still unbroken. What are the dominant 1-loop contributions for 
other scalar particles, say the stop ? Compute the processes 
$\mu~ \to~ e + ~\gamma$ and $K^0 - \bar{K}^0$ mixing in this limit. 

\item As we have seen, W is a holomorphic function and that there
are two Higgs doublets giving masses to up type and down type
quarks separately. (a) Give examples of operators which are gauge 
invariant but  non-holomorphic ?  (b) Show that such operators 
involving the Higgs fields will lead to Yukawa like couplings with
the ``wrong" Higgs. Study the implications of such couplings.  

\end{itemize}
\noindent
\textit{Historical Note} \\
Supersymmetries were first introduced in the context of string 
theories by Ramond. In quantum field theories,
this symmetry is realised through fermionic generators, thus
escaping the no-go theorems of Coleman and Mandula \cite{weinbergv3}. 
The simplest Lagrangian realising this symmetry in four dimensions 
was built by  Wess and  Zumino which contains a spin ${1 \over 2}$ fermion 
and a scalar. 

In particle physics, supersymmetry plays an important role in protecting
the Higgs mass.  To understand how it protects the Higgs mass, let us consider
the hierarchy problem once again. The Higgs mass enters as a bare mass
parameter in the Standard Model lagrangian, eq.(\ref{lscalar}). 
Contributions from the self energy diagrams of the Higgs are 
quadratically divergent pushing the Higgs mass up to cut-off scale. 
In the absence of any new physics at the intermediate 
energies, the cut-off scale is typically $M_{GUT}$ or $M_{planck}$.  
Cancellation of these divergences with the bare mass parameter 
would require fine-tuning of order one part in $10^{-36}$ rendering 
the theory `unnatural'\cite{natural}. In a complete GUT model 
like SU(5) this might reflect as a severe problem of doublet-triplet 
splitting \cite{buras,gildener}. 
On the other hand, if one has 
additional contributions, say, for example, for the diagram with the
Higgs self coupling, there is an additional contribution from a fermionic 
loop, with the fermion carrying 
the same mass as the scalar, the contribution from this additional diagram 
would now cancel the quadratically divergent part of the SM diagram, 
with the total contribution now
being only logarithmically divergent. If this mechanism needs to work for
all the diagrams, not just for the Higgs self-coupling and for all orders
in perturbation theory,  it would require
a symmetry which would relate a fermion and a boson with same mass.
Supersymmetry is such a symmetry.

\subsection{R-parity}

In the previous section, we have seen that there are terms in the superpotential,
eq.(\ref{superrpv}) which are invariant under the Standard Model gauge
group $G_{SM}$ but however violate baryon ($\mathrm{B}$) and individual
lepton numbers ($\mathrm{L}_{e,\mu,\tau}$). At the first sight, 
it is bit surprising :  the
matter \textit{superfields} carry the same quantum numbers under the $G_{SM}$ 
just like the ordinary matter fields  do in the Standard Model and $\mathrm{B}$ 
and $\mathrm{L}_{e,\mu,\tau}$ violating terms are not present in the Standard 
Model. The reason can be traced to the fact that in the MSSM, where matter 
sector is represented in terms of  superfields, 
there is no distinction between the fermions and the bosons of the model. 
 In the Standard Model, the Higgs field is a boson and the leptons
and quarks are fermions and they are different representations of
the Lorentz group. This distinction is lost in the MSSM,
the Higgs superfield, $H_1$ and the lepton superfields 
$L_i$ have the same quantum numbers under $G_{SM}$ and given that
they are both (chiral) superfields, there is no way of distinguishing
them.  For this reason, the second part of the superpotential $W_2$
makes an appearance in supersymmetric version of the Standard Model. 
In fact, the first three terms of eq.(\ref{superrpv}) can be achieved 
by replacing $H_1 ~\to~ L_i$ in the terms containing $H_1$ of $W_1$.

The first three terms of the second part of the superpotential $W_2$
(eq.(\ref{superrpv})), are lepton number violating whereas the last term 
is baryon number violating. The simultaneous presence of both these 
interactions can lead to proton decay, for example, through a squark exchange. 
An example of such an process in given in Figure 1. Experimentally the
proton is quite stable. In fact its life time is pretty large 
$\ger~ \mathcal{O}(10^{33})$ years \cite{skproton}. Thus products of
these couplings ($\lambda''$ and  one of ($\lambda'~ ,\epsilon,~\lambda$)
which can lead to proton decay are severely constrained to be of the
order of $\mathcal(O)(10^{-20})$\footnote{The magnitude of these constraints
depends also on the scale of supersymmetry breaking, which we will come
to discuss only in the next section. For a list of constraints on R-violating
couplings, please see G. Bhattacharyya \cite{gautamB}.}. 
Thus to make the MSSM phenomenologically
viable one should expect these couplings in $W_2$ to take such extremely 
small values. 

\begin{center}
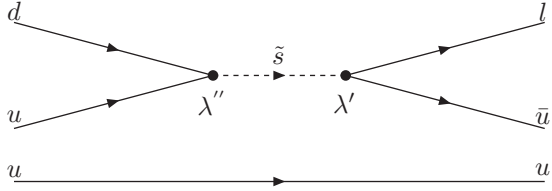
\begin{figure}
\begin{picture}(300,200)(0,0)
\ArrowLine(0,50)(200,50) \Text(0,54)[]{$u$} \Text(200,55)[]{$u$}
\ArrowLine(0,70)(75,90)  \Text(0,75)[]{$u$}
\ArrowLine(0,110)(75,90) \Text(0,115)[]{$d$} \Vertex(75,90){2}
\Text(75,80)[]{$\lambda^{''}$}
\DashArrowLine(75,90)(125,90){2} \Vertex(125,90){2} \Text(100,98)[]{$\tilde{s}$}
\Text(125,80)[]{$\lambda'$}
\ArrowLine(125,90)(200,110) \Text(200,115)[]{$\bar{l}$}
\ArrowLine(125,90)(200,70)  \Text(200,75)[]{$\bar{u}$}
\end{picture}
\caption{A sample diagram showing the decay of the proton in the presence
of R-parity violating couplings.}
\end{figure}
\end{center}

A more natural way of dealing with such small numbers for these couplings
would be to set them to be zero. This can be
arrived at by imposing a discrete symmetry on the
lagrangian called R-parity. R-parity  has been originally
introduced as a discrete R-symmetry \footnote{R-symmetries are symmetries
under which the $\th$ parameter transform non-trivially.} by Ferrar and
Fayet \cite{fayetfarrar} and then later realised to be of the following 
form by Ferrar and Weinberg \cite{weinfarrar} acting on the component fields:
\be
R_p = (-1)^{3 (B-L) + 2s},
\ee
where  B and L represent the Baryon and Lepton number respectively and
s represents the spin of the particle. Under R-parity the transformation 
properties of various superfields can be summarised as:

\bea
\label{rptrans}
\{ V_s^A , V_w^I , V_y \} &\row& \{ V_s^A , V_w^I , V_y \} \nonumber \\
\th &\row& - \th^{\star} \nonumber \\
\{Q_i, U^c_i, D^c_i, L_i, E_i^c \}&\row&
 -\{Q_i,U^c_i,D_i^c,L_i,E_i^c \}\nonumber \\
\{ H_1, H_2\}&\row& \{H_1,H_2\}
\eea
Imposing these constraints on the superfields will now set 
all the couplings in $W_2$ to zero. 

Imposing R-parity has an advantage that it provides a natural
candidate for dark matter. This can be seen by observing that
R-parity distinguishes a particle from its superpartner.  This ensures
that every interaction vertex has at least two supersymmetric partners
when R-parity is conserved. 
The lightest supersymmetric particle (LSP) cannot decay  in to a pair
of SM particles and remains
stable. R-parity can also be thought of as a remnant 
symmetry theories with an additional $U(1)$ symmetry, which is
natural in  a large class of supersymmetric Grand Unified theories.
Finally, one curious fact about R-parity : it should be noted
that  R-parity constraints baryon and lepton number violating 
couplings of dimension four or rather only at the renormalisable
level. If one allows for non-renormalisable operators in the 
MSSM, \textit{i.e} that is terms of dimension more than three in
the superpotential, they can induce dim 6 operators which 
violate baryon and lepton numbers  at the lagrangian level and
are still allowed by R-parity.  Such operators are typically suppressed
by high mass scale $\sim M_{Pl} $ or $M_{GUT}$ and thus are less dangerous.
In the present  set of lectures,  we will always impose R-parity in the MSSM 
so that the proton does not decay, though there are alternatives
to R-parity which can also make proton stable.

\subsubsection{\textit{Think it over }}
\begin{itemize}
\item Is imposing R-parity the only way to get rid of the terms 
which lead to proton decay ? (Hint: For proton decay to occur
both $\mathrm{L}$ and $\mathrm{B}$ violating operators are 
required.  R-parity removes both these sets of operators which
is unnecessary. We can think of discrete symmetries which can
remove only either $\mathrm{B}$ or $\mathrm{L}$ type of operators.)
See for example\cite{rosshall}.
\end{itemize}

\section{STEP 3: Supersymmetry breaking}
So far, we have seen that the Supersymmetric Standard Model lagrangian 
can also be organised in a similar way like the Standard Model lagrangian
though one uses functions of superfields now to get the lagrangian rather
than the ordinary fields. In the present section we will cover the last
part (term) of the total MSSM lagrangian 
\begin{equation}
\mathcal{L}_{\mbox{MSSM}} = 
\mathcal{L}_{\mbox{gauge/kinetic}} \left( K(\Phi, V) \right)  + 
\mathcal{L}_{\mbox{yukawa}} \left(W(\Phi) \right) + 
\mathcal{L}_{\mbox{scalar}} \left(F^2, D^2 \right)+ 
\mathcal{L}_{\mbox{SSB}} 
\end{equation}
which we have left out so far and that concerns supersymmetry breaking (SSB).
Note that the first three terms are essentially from 
$\mathcal{L}_{\mbox{MSSM}}^{(0)}$ of eq.(\ref{mssmlag}). 
In Nature, we do not observe supersymmetry. Supersymmetry breaking 
has to be incorporated in the MSSM to make it realistic. 
In a general lagrangian,  supersymmetry can be broken 
spontaneously if the auxiliary fields F or D appearing in the definitions 
of the chiral and vector superfields respectively attain a vacuum expectation
value ({\it vev}). If the $F$ fields get a \textit{vev}, it is called
$F$-breaking whereas if the $D$ fields get a \textit{vev}, it is called
$D$-breaking. 

Incorporation of spontaneous SUSY breaking in MSSM would mean that
at least one (or more) of the F-components  corresponding to one ( or 
more) of the MSSM chiral (matter) superfields would attain a vacuum
expectation value. However, this approach fails as this leads to 
phenomenologically unacceptable prediction that at least one of the
super-partner should be lighter (in mass) than the ordinary particle. 
This is not valid phenomenologically as such a light super partner
(of SM particle) has been ruled out experimentally. One has to think
of a different  approach for incorporating supersymmetry breaking
in to the MSSM \cite{Luty}.

\begin{figure}[t]
\includegraphics[width=0.88\textwidth]{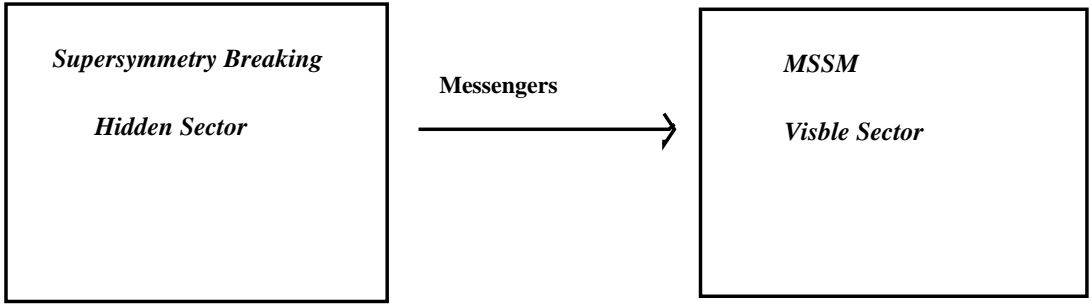}
\caption{A schematic diagram showing SUSY breaking using Hidden sector models}
\end{figure}

One of the most popular and successful approaches has been to assume
another sector of the theory consisting of superfields which are not
charged under the Standard Model gauge group. Such a sector of the
theory is called `Hidden Sector' as they cannot been "seen" like
the Standard Model particles and remain hidden.  Supersymmetry can be
broken spontaneously in this sector.  This information is communicated
to the visible sector or MSSM through a messenger sector. 
The messenger sector can
be made up of gravitational interactions or ordinary gauge interactions.  
The communication of supersymmetry breaking leads to  supersymmetry
breaking terms in MSSM. 
 Thus, supersymmetry is not broken
spontaneously within the MSSM, but \textit{explicitly} by adding
supersymmetry breaking terms in the lagrangian. 

However, not all supersymmetric terms can be added. We need to add
only those terms which do not re-introduce quadratic divergences back
into the theory\footnote{Interaction terms and other couplings which do
not lead to quadratically divergent (in cut-off $\Lambda$) terms in the 
theory once loop corrections are taken in to consideration. It 
essentially means we only add dimensional full couplings which are 
supersymmetry breaking.}. 
It should be noted that in most models of spontaneous supersymmetry breaking, 
only such terms are generated. These terms which are called ``soft" 
supersymmetry breaking terms can be classified as follows:
\begin{itemize}
\item a) Mass terms for the gauginos which are a part of the various
vector superfields of the MSSM. 
\item b) Mass terms for the scalar particles, 
$m^2_{\phi_{ij}} ~\phi_i^\star \phi_j$ with $\phi_{i,j}$ representing 
the scalar partners of chiral superfields of the MSSM. 
\item c) Trilinear scalar interactions, $A_{ijk} \phi_i \phi_j \phi_k$
	corresponding to  the cubic terms in the superpotential. 
\item d) Bilinear scalar interactions, $B_{ij} \phi_i \phi_j$ corresponding
	to the bilinear terms in the superpotential. 
\end{itemize}
Note that all the above terms are dimensionful. Adding these terms would 
make the MSSM non-supersymmetric and thus realistic.  The total MSSM 
lagrangian is thus equal to 
\be
\lm_{total} = \lm_{MSSM}^{(0)} + \lm_{SSB}
\ee
with $\lm_{MSSM}^{(0)}$ given in eq.(\ref{mssmlag}). Sometimes in literature
we have $ \lm_{SSB} = \lm_{soft}$. Let us now see the complete
list of all the soft SUSY breaking terms   one can incorporate in the MSSM: 

\begin{enumerate}

\item \textit{Gaugino Mass terms}: Corresponding to the three vector
superfields  (for gauge groups $U(1)$, $SU(2)$ and $SU(3)$) we 
have  $\tilde{B}, \tilde{W}$ and $\tilde{G}$) we have
three gaugino mass terms which are  given as  $M_1 \tilde{B} \tilde{B}$,
$M_2 \tilde{W}_I \tilde{W}_I $ and $M_3 \tilde{G}_A \tilde{G}_A$, where
$I(A)$ runs over all the $SU(2)(SU(3))$ group generators. 

\item \textit{Scalar Mass terms}: For every scalar in each 
chiral (matter) superfield , we can add a mass term of the form
$m^2~ \phi_i^\star \phi_j $. Note that the generation indices $i,j$ 
\textit{need} not be the same. Thus the mass terms can violate 
flavour. Further, given that SUSY is broken prior to 
$SU(2)\times U(1)$ breaking , all these mass terms for the scalar
fields should be written in terms of their `unbroken' 
$SU(2) \times U(1)$ representations. Thus the scalar mass terms
are : $m_{Q_{ij}}^2  \tilde{Q}_i^\dagger \tilde{Q}_j$ ,
$m_{u_{ij}}^2  \tilde{u^c}_i^\star \tilde{u^c}_j$ ,
$m_{d_{ij}}^2  \tilde{d^c}_i^\star \tilde{d^c}_j$ ,
$m_{L_{ij}}^2  \tilde{L}_i^\dagger \tilde{L}_j$ ,
$m_{e_{ij}}^2  \tilde{e^c}_i^\star \tilde{e^c}_j$ ,
 $m_{H_1}^2  H_1^\dagger H_1$ and $m_{H_2}^2  H_2^\dagger H_2$.

\item \textit{Trilinear Scalar Couplings}: 
As mentioned again, there are only three types of trilinear
scalar couplings one can write which are $G_{SM}$ gauge invariant. 
In fact, their form exactly follows from the  Yukawa couplings.
These are : $A^u_{ij} \tilde{Q}_i \tilde{u}^c_j H_2 $,
 $A^d_{ij} \tilde{Q}_i \tilde{d}^c_j H_1 $ and 
 $A^e_{ij} \tilde{L}_i \tilde{e}^c_j H_1 $.

\item \textit{Bilinear Scalar Couplings}: 
Finally, there is only one bilinear scalar coupling (other than the
mass terms) which is gauge invariant. The form of this term also follows
from the superpotential. It is given as : $B H_1 H_2 $. 

\end{enumerate}
\noindent
Adding all these terms completes the lagrangian for the MSSM. However, 
the particles are still not in their `physical' basis as 
$SU(2) \times U(1)$ breaking is not yet incorporated. Once incorporated
the physical states of the MSSM and their couplings could be derived. 

\section{STEP 4: $SU(2) \times U(1)$ breaking}
As a starting point, it is important to realize that the 
 MSSM is a two Higgs doublet model \textit{i.e}, SM with two Higgs doublets
instead of one, with a different set of couplings~\cite{higgshunter}.  
Just as in Standard Model,
spontaneous breaking of  $SU(2)_L \times U(1)_Y ~\to~ U(1)_{EM}$ 
can be incorporated here too. Doing this leads
to constraints  relating various parameters of the model. To see this, 
consider the neutral Higgs part of the total scalar potential including
the soft terms. It is given as 

\bea
\label{vhneutral}
V_{scalar}&=& (m_{H_1}^2 + \m^2) |H_1^0|^2 + (m_{H_2}^2 + \m^2 ) |H_2^0|^2 -
(B_\m \m H_1^0 H_2^0 + H.c) \nonumber \\
&+&  {1 \over 8} ( g^2 + g^{\prime 2}) ({H_2^0}^2 - {H_1^0}^2)^2 + \ldots,
\eea

\noindent
where $H_1^0, H_2^0$ stand for the neutral Higgs scalars and we have 
parameterised the bilinear soft term $B \equiv B_\m \m$.
Firstly, we should require that the potential should be bounded 
from below. This gives the condition (in field configurations where 
the D-term goes to zero, \textit{i.e}, the second line in eq.(\ref{vhneutral})): 
\be
2 B_\mu < 2 |\mu|^2~ +~ m_{H_2}^2~ + m_{H_1}^2 
\ee 
Secondly, the existence of a minima for the above potential would require 
at least one of the Higgs mass squared to be negative giving the condition,
(determinant of the $2 \times 2$ neutral Higgs mass squared matrix should be
negative)
\be
 B_\mu^2 >  ( |\mu|^2~ +~ m_{H_2}^2~)~(|\mu|^2 + m_{H_1}^2)
\ee 
In addition to ensuring the existence of a minima, one would also 
require that the minima should be able to reproduce the standard model 
relations i.e, correct gauge boson masses. We insist that both
the neutral Higgs attain vacuum expectation values :
\be
\label{vevs}
<H^0_1> = {v_1 \over \sqrt{2}} \;\;\;\;; \;\;\; <H_2^0> = {v_2 \over \sqrt{2}} 
\ee 
and furthermore we define $$ ~v_1^2 + v_2^2~ = ~v^2 ~= ~246^2~ \mbox{GeV}^2,$$
where $v$ represents the vev of the Standard Model (SM) Higgs field. 
However, these \textit{vev}s should correspond to the minima 
of the MSSM potential.  The minima  are derived by requiring  
$\partial V / \partial H_1^0~ =~ 0 $
and  $\partial V / H_2^0~ =~ 0 $ at the minimum, where the form of 
$V$ is given in eq.(\ref{vhneutral}).  These derivative conditions lead
to relations between the various parameters of the model at the minimum
of the potential. We have, using the  Higgs vev (\ref{vevs})
and the formulae for\footnote{In this lecture note, we will be using
$g_2~ = g~ =~ g_W $ for the SU(2) coupling, whereas $g' = g_1 $ for
the $U(1)_Y$ coupling and $g_s = g_3$ for the SU(3) strong coupling.}
$M_Z^2~ =~ {1 \over 4} (g^2 + g^{'~2}) v^2 $, the minimisation conditions 
can rewritten as 
\bea
\label{minime}
{1 \over 2} M_Z^2 &=& {m_{H_1}^2 - \tan^2 \b~ m_{H_2}^2 \over
\tan^2 \b  - 1} - \m^2  \nonumber \\
\mbox{Sin} 2 \b &=& { 2 B_\m ~\m~ \over m_{H_2}^2 + m_{H_1}^2 + 2 \m^2 },
\eea
where we have used the definition  $\tan \b = v_2 /v_1 $ as the ratio of 
the vacuum expectation values of $H_2^0$ and $H_1^0$ respectively.
Note that the parameters $m_{H_1}^2,~ m_{H_2}^2 , B_\mu$ are all supersymmetry
breaking `soft' terms. $\mu$ is the coupling which comes
in the superpotential giving the supersymmetry conserving masses to the
Higgs scalars. These are related to the Standard Model parameters $M_Z$ and
a ratio of \textit{vev}s, parameterised by an angle tan$\beta$. Thus these
conditions relate SUSY breaking soft parameters with the SUSY conserving
ones and the Standard Model parameters. For any
model of supersymmetry to make contact with reality, the above two conditions
(\ref{minime} )need to be satisfied. 

The above minimisation conditions are given for the `tree level' potential only.
1-loop corrections a {'la} Coleman-Weinberg  can significantly modify these 
minima. We will discuss a part of them in later sections when we discuss
the Higgs spectrum. Finally we should mention that, in a more concrete approach,
one should consider the entire scalar potential including all the scalars in
the theory, not just confining ourselves to the neutral Higgs scalars. For such
a potential one should further demand that there are no deeper minima which
are color and charge breaking (which effectively means none of the colored
and charged scalar fields get vacuum expectation values). These conditions
lead to additional constraints on parameters of the MSSM\cite{casasdimo}.

\subsubsection{\textit{Think it over }}
\begin{itemize}
\item In the MSSM, we have considered here contains two Higgs doublets. 
In addition to $H_1$ and $H_2$, consider an additional Higgs field field $S$, 
which transforms as a singlet under all the gauge groups of $G_{SM}$. 
Write down the superpotential
including the singlet field $S$ invariant under $G_{SM}$. Derive the 
corresponding scalar potential
including the soft SUSY breaking terms. Minimise the neutral Higgs
potential and derive the electro-weak minimisation
conditions. How many are there and what are they? (Hint: Assume the $S$
field also develops a \textit{vev} and that its \textit{vev} is much 
larger than $v_1$ and $v_2$. )
\end{itemize}

\section{Step 5: Mass spectrum}
We have seen in the earlier section, supersymmetry breaking terms introduce
mass-splittings between ordinary particles and their super-partners. Given
that particles have zero masses in the limit of exact $G_{SM}$, only 
superpartners are given soft mass terms. After the $SU(2)~\times U(1)$ 
breaking, ordinary particles as well as superparticles attain mass terms.
For the supersymmetric partners, these mass terms are  either
additional contributions or  mixing
terms between the various super-particles. Thus, just like in the case
of ordinary SM fermions, where one has to diagonalise the fermion mass
matrices to write the lagrangian in the `on-shell' format or the physical
basis, a similar diagonalisation has to be done for the
super-symmetric particles and their mass matrices. 

\subsubsection{The Neutralino Sector}
To begin with lets start with the gauge sector. The superpartners of the 
neutral gauge bosons (neutral gauginos)  and the fermionic partners of 
the neutral higgs bosons (neutral higgsinos) mix to form Neutralinos. The neutralino
mass matrix in the basis 
$$ \mathcal{L}~ \supset~ {1 \over 2}~ \Psi_N \mathcal{M}_N \Psi_N^T~ + H.c$$ 
where $\Psi_N = \{\tilde B,~\tilde W^0,\tilde H_1^0,\tilde H_2^0\}$
is given as :
\be
\mathcal{M}_N ~=~
\left(\matrix{M_1&0 & - M_Z c \beta~ s \theta_W& M_Z s \beta~ s \theta_W
\cr0& M_2&  M_Z c \beta~ c \theta_W& M_Z s \beta~ c \theta_W \cr
- M_Z c \beta~ s \theta_W& M_Z c \beta~ c \theta_W &0& -\mu\cr
M_Z s \beta~ s \theta_W& -M_Z s \beta ~c \theta_W & -\mu&0} \right),
\ee
with $c \beta (s \beta)$ and $c \theta_W (s \theta_W)$ standing for $\cos \beta(\sin  \beta)$
and $\cos \theta_W(\sin \theta_W) $ respectively. As mentioned earlier, $M_1$ and $M_2$
are the soft parameters, whereas $\mu$ is the superpotential parameter, thus
SUSY conserving. The angle $\beta$ is typically taken as a input parameter,
$tan\beta = {v_2 /v_1}$ whereas $\theta_W$ is the Weinberg angle given by
the inverse tangent of the ratio of the gauge couplings as in the SM.
 Note that the neutralino mass matrix
being a Majorana mass matrix is complex symmetric in nature. Hence it is diagonalised 
by a unitary matrix $N$,
\bea
N^* \cdot M_{\tilde N}
\cdot N^\dagger = \mbox{Diag.}
(m_{\chi_1^0}, m_{\chi_2^0}, m_{\chi_3^0}, m_{\chi_4^0})
\eea 
The states are rotated by $\chi_i^0 = N^\star \Psi$ to go the physical basis.
\subsubsection{The Chargino Sector}
In a similar manner to the neutralino sector, all the 
fermionic partners of the charged gauge bosons and of the 
charged  Higgs bosons mix after electroweak symmetry breaking.
However, they combine in a such a way that a Wino-Higgsino
Weyl fermion pair forms a Dirac fermion called the chargino. This mass matrix
is given as 
\be
\mathcal{L} \supset -\frac{1}{2} \left(\matrix{\tilde W^- & \tilde H_1^-} \right) \;
\left(\matrix{M_2  & \sqrt{2} M_W \sin \beta \cr \sqrt{2} M_W \cos
\beta & \mu} \right) \left(\matrix{\tilde W^+ \cr \tilde H_2^+
} \right),
\ee
Given the  non-symmetric (non-hermitian) matrix  nature of this
matrix, it is diagonalised by a bi-unitary transformation, $U^* \cdot M_C \cdot 
V^\dagger = \mbox{Diag.}(m_{\chi_1^+},m_{\chi_2^+})$. The chargino
eigenstates are typically represented by $\chi^\pm$ with mass eigenvalues
$m_{\chi^\pm}$. The explicit forms for $U$ and $V$ can be found by the 
eigenvectors of $M_C M_C^\dagger$ and $M_C^\dagger M_C$ respectively \cite{haberkane}. 
\subsubsection{The Sfermion Sector}
Next let us come to the sfermion sector. Remember that we have added
different scalar fields for the right and left handed fermions in the
Standard Model. After electroweak symmetry breaking, the sfermions
 corresponding to the left fermion and the right fermion mix with each
 other.  Furthermore while writing
down the mass matrix for the sfermions, we should remember 
that these terms could break the flavour \textit{i.e}, we can have
mass terms which mix different generation. Thus, in  general
the sfermion mass matrix is a   $6 \times 6$ mass  matrix given
as : $$ \xi^\dagger ~M_{\tilde f}^2 \xi~~ ; ~~~ \xi =\{{\tilde f_{L_i}},{\tilde f_{R_i}}\} $$ 
From the total  scalar potential, the mass matrix for these sfermions can be derived
using standard definition given as
\be
\label{stdscalarmass}
m_{ij}^2 = \left( \begin{array}{cc}  {\partial^2 V \over \partial \phi_i \partial \phi_j^\star} &
  {\partial^2 V \over \partial \phi_i \partial \phi_j }\\
 {\partial^2 V \over \partial \phi_i^\star \partial \phi_j^\star} &
  {\partial^2 V \over \partial \phi_i^\star \partial \phi_j } \end{array} \right)
\ee 
Using this for sfermions, we have :
\be
M_{\tilde f}^2 \; = \;
\left(\matrix{m_{\tilde f_\LL}^2 & m_{\tilde f_\LR}^2 \cr {m_{\tilde f_\LR}^{2\,\dagger}}
&m_{\tilde f_\RR}^2 } \right),
\ee where each of the above entries represents $3 \times 3$ matrices in the
generation space. More specifically, they have the form (as usual, $i,j$ are
generation indices):
\bea
m^2_{\tilde f_{L_i L_j}} &= &  M^2_{\tilde f_{L_i L_j}} + m^2_f \delta_{ij} 
+ M_Z^2 \cos 2 \beta (T_3 + \sin^2 \theta_W Q_\elm) \delta_{ij} \nonumber \\
m_{{\tilde f}_{L_i R_j}}^2& =& \left (\big(  Y^A_f\cdot ^{v_2}_{v_1} - m_f \mu ^{\tan \beta}_{\cot \beta}\big) \;\; \mbox{for}\; f=^{e,d}_u \right) \delta_{ij}  \nonumber \\[.2cm] 
m_{\tilde f_\RR}^2 &=&  M^2_{\tilde f_{R_{ij}}} + \left( m^2_f + M_Z^2 \cos 2 \beta
\sin^2 \theta_W Q_\elm \right) \delta_{ij}
\eea
In the above, $M^2_{\tilde f_{L}}$ represents the soft mass term for the corresponding 
fermion ($L$ for left, $R$ for right), $T_3$ is the eigenvalue of the diagonal
generator of $SU(2)$,  $m_f$ is the mass of the fermion with 
$Y$ and $Q_{em}$ representing the hypercharge and electromagnetic charge (in units of
the charge of the electron ) respectively. 
The  sfermion mass matrices are hermitian  and are thus diagonalised by a unitary rotation, $R_{\tilde{f}} R_{\tilde{f}} ^\dagger = 1 $:
\be
R_{\tilde f} \cdot M_{\tilde f} \cdot R_{\tilde f}^\dagger = 
\mbox{Diag.}(m_{\tilde f_1},m_{\tilde f_2},\dots,m_{\tilde f_6})
\ee 

\subsubsection{The Higgs sector}
Now let us turn our attention to the Higgs fields.
We will use again use the standard formula of eq.(\ref{stdscalarmass}), to derive the
Higgs mass matrices.  The eight  Higgs degrees of freedom form a $8 \times 8$ Higgs
mass matrix which breaks down diagonally in to three  $ 2 \times 2 $ mass 
matrices\footnote{The discussion in this section closely follows from the discussion presented 
in Ref.\cite{rohinibook}}.

The mass matrices are divided in to charged sector, CP odd neutral and CP even neutral. 
This helps us in identifying the goldstone modes and the physical spectrum in an simple manner.
 Before writing down the mass matrices, let us first define the
following parameters : $$ m_1^2 = m_{H_1}^2 + \mu^2, \;\;\;
m_2^2 = m_{H_2}^2 + \mu^2 , \;\;\; m_3^2 =  B_\mu \mu. $$ 
\noindent
In terms of these parameters, the various mass matrices and the corresponding physical
states obtained after diagonalising the mass matrices are given below: \\[4pt]
\textit{Charged Higgs and Goldstone Modes}: \\
\be
\left( \begin{array}{cc} H_1^+ & H_2^+ \end{array} \right) 
\left( \begin{array}{cc} 
m_1^2 + {1 \over 8} (g_1^2 + g_2^2) (v_1^2 - v_2^2) + {1 \over 4} g_2^2 v_2^2 & 
m_3^2 + {1 \over 4} g_2^2 v_1 v_2 \\ 
m_3^2 + {1 \over 4} g_2^2 v_1 v_2  &
m_2^2 - {1 \over 8} (g_1^2 + g_2^2) (v_1^2 - v_2^2) + {1 \over 4} g_2^2 v_2^2 
\end{array} \right) 
\left( \begin{array}{c} H_1^- \\ H_2^- \end{array} \right) 
\ee
Using the minimisation conditions (\ref{minime}), this matrix becomes,
\be
\left( \begin{array}{cc} H_1^+ & H_2^+ \end{array} \right) 
({m_3^2 \over v_1 v_2} + {1 \over 4} g_2^2) 
\left( \begin{array}{cc} 
v_2^2 & v_1 v_2 \\ 
v_1 v_2 & v_1^2  \\ 
\end{array} \right) 
\left( \begin{array}{c} H_1^- \\ H_2^- \end{array} \right) 
\ee
which has determinant zero leading to the two eigenvalues as :
\begin{eqnarray}
m_{G^\pm}^2& =& 0 \nonumber \\
m_{H^\pm}^2& =& \left({m_3^2 \over v_1 v_2} + {1 \over 4} g_2^2\right) (v_1^2 + v_2^2), \\
& =& {2 m_3^2 \over sin 2 \beta } + M_W^2 
\end{eqnarray}
where $G^\pm$ represents the Goldstone mode. The physical states are obtained
just by rotating the original states in terms of the $H_1,~ H_2$ fields by an 
mixing angle.  The mixing angle in the present case (in the unitary gauge) 
is just tan$\beta$: 
\be
\left( \begin{array}{c} H^\pm \\ G^\pm \end{array} \right) 
= \left( \begin{array}{cc} 
sin\beta & cos\beta \\
-cos\beta & sin\beta
\end{array} \right) 
\left( \begin{array}{c} H^\pm \\ G^\pm \end{array} \right) 
\ee
\noindent
\textit{CP odd Higgs and Goldstone Modes}: \\[2pt]
Let us now turn our attention to the CP-odd Higgs sector. The mass matrices
can be written in a similar manner but this time for imaginary components
of the neutral Higgs. 
\be
\left( \begin{array}{cc} Im H_1^0 & Im H_2^0 \end{array} \right) 
\left( \begin{array}{cc} 
m_1^2 + {1 \over 8} (g_1^2 + g_2^2) (v_1^2 - v_2^2)  & 
m_3^2 \\ 
m_3^2  &
m_2^2 - {1 \over 8} (g_1^2 + g_2^2) (v_1^2 - v_2^2)  
\end{array} \right) 
\left( \begin{array}{c} Im H_1^0 \\ Im H_2^0 \end{array} \right) 
\ee
As before, again using the minimisation conditions, this matrix becomes,
\be
\left( \begin{array}{cc} Im H_1^0 & Im H_2^0 \end{array} \right) 
m_3^2
\left( \begin{array}{cc} 
v_2/ v_1  & 1 \\ 
1 & v_1/v_2 \\ 
\end{array} \right) 
\left( \begin{array}{c} Im H_1^0 \\ Im H_2^0 \end{array} \right) 
\ee
which has determinant zero leading to the two eigenvalues as :
\begin{eqnarray}
m_{G^0}^2& =& 0 \nonumber \\
\label{pseudohiggs}
m_{A^0}^2& =& \left({m_3^2 \over v_1 v_2} \right) (v_1^2 + v_2^2)~~ =~~ {2 m_3^2 \over sin 2 \beta}
\end{eqnarray}
Similar to the charged sector, the mixing angle between these two states 
in the unitary gauge is again just tan$\beta$. 
\be
{1 \over \sqrt{2}} 
\left( \begin{array}{c} A^0 \\ G^0 \end{array} \right) 
= 
\left( \begin{array}{cc} 
sin\beta & cos\beta \\
-cos\beta & sin\beta
\end{array} \right) 
\left( \begin{array}{c} Im H^0_1 \\ Im H_2^0\end{array} \right) 
\ee
\noindent
\textit{CP even Higgs}: \\
Finally, let us come to the real part of the neutral Higgs sector. The mass matrix 
in this case is given by the following. 
\be
\left( \begin{array}{cc} Re H_1^0 & Re H_2^0 \end{array} \right) 
~{1 \over 2}~ 
\left( \begin{array}{cc} 
2 m_1^2 + {1 \over 4} (g_1^2 + g_2^2) (3 v_1^2 - v_2^2)  & 
-2 m_3^2 - {1 \over 4} v_1 v_2 (g_1^2 + g_2^2)  \\ 
-2 m_3^2 - {1 \over 4} v_1 v_2 (g_1^2 + g_2^2) & 
2 m_2^2 + {1 \over 4} (g_1^2 + g_2^2) (3 v_2^2 - v_1^2)  
\end{array} \right) 
\left( \begin{array}{c} Re H_1^0 \\ Re H_2^0 \end{array} \right) 
\ee
Note that in the present case, there is no Goldstone mode. As
before, we will use the minimisation conditions and further
using the definition of $m_A^2$ from eq.(\ref{pseudohiggs}),
we have :
\be
\label{cpevenhiggs}
\left( \begin{array}{cc} Re H_1^0 & Re H_2^0 \end{array} \right) 
\left( \begin{array}{cc} 
m_A^2 sin^2 \beta + M_z^2 cos \beta  & 
- (m_A^2 + m_Z^2) sin\beta cos\beta  \\ 
- (m_A^2 + m_Z^2) sin\beta cos\beta  &
m_A^2 cos^2 \beta + M_z^2 sin \beta  
\end{array} \right) 
\left( \begin{array}{c} Re H_1^0 \\ Re H_2^0 \end{array} \right) 
\ee
The  matrix has two eigenvalues which are given by the two signs 
of the following equation:
\be
m^2_{H,h} = {1 \over 2} \left[ m_A^2+ m_Z^2 \pm \{ (m_A^2 + m_Z^2)^2 - 4 m_Z^2 m_A^2 cos^2 2 \beta\}^{1/2} \right]
\ee 
The heavier eigenvalue $m_H^2$, is obtained by taken the positive sign, whereas the lighter
eigenvalue $m_h^2$ is obtained by taking the negative sign respectively. The mixing angle
between these two states can be read out from the mass matrix of the above\footnote{The mixing
angle for a $ 2 \times 2 $ symmetric matrix, $C_{ij}$ is given by 
$$tan 2 \theta = 2 C_{12} /(C_{22} - C_{11}).$$} as : 
\be
tan~ 2\alpha = {m_A^2 + m_Z^2  \over m_A^2 - m_Z^2} ~tan ~ 2 \beta 
\ee 
\noindent
\textit{Tree Level Catastrophe}: \\
So far we have seen that out of the eight Higgs degrees of freedom, three of them form the
Goldstone modes after incorporating $SU(2) \times U(1) $ breaking and there are five 
\textit{physical} Higgs bosons fields in the MSSM spectrum. These are the charged Higgs ($H^\pm$)
a CP-odd Higgs ($A$) and two CP-even Higgs bosons ($h,H$). From the mass spectrum analysis
above, we have seen that the mass eigenvalues of these Higgs bosons are related to each other. 
In fact, putting together all the eigenvalue equations, we  summarise the relations between
them as follows :
\bea
\label{higgsmassrelations}
m^2_{H^\pm} &= &m_A^2 + m_W^2 > \max(M_W^2, m_A^2) \nonumber \\
m_h^2 + m_H^2 &=& m_A^2 + m_Z^2 \nonumber \\
m_H &>& max(m_A,m_Z) \nonumber \\
m_h &<& min(m_A, M_Z)|cos2 \beta| < min (m_A,m_Z) 
\eea 
Let us concentrate on the last relation of the above eq.(\ref{higgsmassrelations}). The
condition on the lightest CP even Higgs mass, $m_h$, tell us that it should be equal to
$m_Z$ in the limit tan$\beta$ is saturated to be maximum, such that 
cos$2\beta ~\rightarrow ~1$ and $m_A~ \rightarrow~ \infty$. If these limits are not 
saturated, it is evident that the light higgs mass is less that $m_Z$. 
This is one of main predictions of MSSM which could make it easily falsifiable from
the current generation of experiments like LEP, Tevatron and the upcoming LHC.
Given that present day experiments have not found a Higgs less that Z-boson mass, it is
tempting to conclude that the MSSM is not realised in Nature. However caution should
be exercised before taking such a route as our results  are valid only at the
tree level. In fact, in a series of papers in the early nineties \cite{oneloophiggs}, 
it has been shown that large one-loop corrections to the Higgs mass can easily circumvent
this limit. 

\noindent
\textit{The light Higgs Spectrum at 1-loop} \\
As mentioned previously, radiative corrections can significantly modify the mass relations 
which we have presented in the previous section. As is evident, these corrections can be
very important for the light Higgs boson mass. Along with the 1-loop corrections previously,
in the recent years dominant parts of two-loop corrections have also been 
available \cite{slavich2loop} with a more complete version recently given\cite{martin2loop}.  
In the following we will present the one-loop corrections to the light Higgs mass and try to
understand the implications for the condition eq.(\ref{higgsmassrelations}). Writing down
the 1-loop corrections to the CP-even part of the Higgs mass matrix as :
\be
M^2_{Re} = M^2_{Re}(0) + \delta M^2_{Re},
\ee
where $M^2_{Re}(0)$ represents the tree level mass matrix given by eq.(\ref{cpevenhiggs}) and
$\delta M^2_{Re}$ represents its one-loop correction. The dominant one-loop correction comes from
the top quark and stop squark loops  which can be written in the following form:
\be
\delta M^2_{Re} = \left( \begin{array}{cc}
\Delta_{11} & \Delta_{12} \\
\Delta_{12} & \Delta_{22} 
\end{array} \right),
\ee
where 
\bea
\Delta_{11} &=& {3 G_F m_t^4 \over 2\sqrt{2} \pi^2 sin^2 \beta} 
\left[ { \mu ( A_t + \mu cot \beta) \over m_{\tilde{t}_1}^2 - m_{\tilde{t}_2}^2} \right]^2 
\left( 2 - {m_{\tilde{t}_1}^2 + m_{\tilde{t}_2}^2 \over m_{\tilde{t}_1}^2 - m_{\tilde{t}_2}^2 }
 ln {m_{\tilde{t}_1}^2 \over 
m_{\tilde{t}_2}^2} \right) \nonumber \\
\Delta_{12} &=& {3 G_F m_t^4 \over 2\sqrt{2} \pi^2 sin^2 \beta} 
\left[ { \mu ( A_t + \mu cot \beta) \over m_{\tilde{t}_1}^2 - m_{\tilde{t}_2}^2} \right]
 ln {m_{\tilde{t}_1}^2 \over m_{\tilde{t}_2}^2} + {A_t \over \mu} \Delta_{11} \nonumber \\
\Delta_{22} &=& {3 G_F m_t^4 \over \sqrt{2} \pi^2 sin^2 \beta} 
\left[  ln {m_{\tilde{t}_1}^2 m_{\tilde{t}_2}^2  \over m_t^2} +
{A_t (A_t +  \mu cot \beta) \over  m_{\tilde{t}_1}^2 -  m_{\tilde{t}_2}} 
 ln {m_{\tilde{t}_1}^2 \over m_{\tilde{t}_2}^2} \right]
+ {A_t \over \mu} \Delta_{11} 
\eea
In the above $G_F$ represents Fermi Decay constant, $m_t$, the top mass, $m_{\tilde{t}_1}^2,~ m_{\tilde{t}_1}^2$
are the eigenvalues of the stop mass matrix and $A_t$ is the trilinear scalar coupling (corresponding to the
top Yukawa coupling) in the stop mass matrix. $\mu$ and the angle $\beta$ have their usual meanings.
Taking in to account these corrections, the condition (\ref{higgsmassrelations}) takes the form:
\be
\label{higgs1loop}
m_h^2~ < ~ m_Z^2 \text{cos}^2 2 \beta + \Delta_{11} \text{cos}^2 \beta + \Delta_{12} \text{sin} 2 \beta + \Delta_{22}
\text{sin}^2 \beta
\ee 
Given that $m_t$ is quite large, almost twice the $m_Z$ mass, for suitable values of the stop masses,
it is clear that the tree level upper limit on the light Higgs mass is now evaded. However, a reasonable
upper limit can still be got by assuming reasonable values for the stop mass. For example assuming 
stop masses to be around 1 TeV and maximal mixing the stop sector, one attains an upper bound on the
light Higgs mass as:
\be
m_h~ \ler~ 135~ \text{GeV}.
\ee
\subsubsection{Feynman Rules}
In this section, we have written down all the mass matrices of the superpartners, their
eigenvalues and finally the eigenvectors which are required to transform the superpartners
in to their physical basis. The feynman rules corresponding to the various vertices have
to be written down in this basis. Thus various soft supersymmetry breaking and supersymmetry 
conserving parameters entering these mass matrices would now determine these couplings 
as well as the masses, which in turn determine the strength of 
various physical processes like crosssections and decay rates. A complete list of the Feynman rules in the mass basis
can be found in various references like Physics Reports like  Haber \& Kane \cite{haberkane} and 
D Chung et. al\cite{Kaneking} and also in textbooks like Sparticles \cite{rohinibook} and
Baer \& Tata \cite{baertata}.  A complete set of Feynman rules is out of reach of this set of
lectures.  Here I will just present two examples to illustrate the points I have been making
here. 
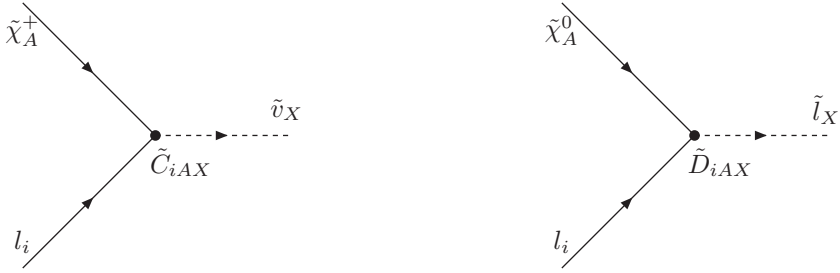
\begin{figure}
\label{chargino}
\begin{tabular}{cc}
\begin{picture}(200,150)(0,0)
\ArrowLine(0,0)(50,50)  \Text(0,10)[]{$l_i$}
\ArrowLine(0,100)(50,50) \Text(0,90)[]{$\tilde{\chi}^+_A$} \Vertex(50,50){2}
\DashArrowLine(50,50)(100,50){2}  \Text(100,60)[]{$\tilde{v}_X$}
\Text(60,40)[]{$\tilde{C}_{iAX}$}
\end{picture} &
\begin{picture}(200,150)(0,0)
\ArrowLine(0,0)(50,50)  \Text(0,10)[]{$l_i$}
\ArrowLine(0,100)(50,50) \Text(0,90)[]{$\tilde{\chi}^0_A$} \Vertex(50,50){2}
\DashArrowLine(50,50)(100,50){2}  \Text(100,60)[]{$\tilde{l}_X$}
\Text(60,40)[]{$\tilde{D}_{iAX}$}
\end{picture}
\end{tabular}
\caption{lepton-slepton-chargino and lepton-slepton-neutralino vertices. }
\end{figure}

Due to the mixing between the fermionic partners of the gauge bosons and the fermionic partners
of the Higgs bosons, the gauge and the yukawa vertices get mixed in MSSM. We will present here
the vertices of fermion-sfermion-chargino and fermion-sfermion-neutralino where this is 
evident. These are presented in Figure 3. \\
\noindent
(i) Fermion-Sfermion-Chargino :\\
This is the first vertex on the left of the figure. The explicit structure of this vertex
is given by:
\be
\tilde{C}_{iAX} = C^R_{iAX} P_R + C^L_{iAX} P_L 
\ee
where $P_L (P_R)$ are the project operators\footnote{$P_L = ( 1 - \gamma_5)/2 $ and
$P_R = ( 1 + \gamma_5)/2 $.} and $C^R$ and $C^L$ are given by
\bea
c^R_{iAX } &=& -g_2 (U)_{A1} R^\nu_{Xi} \\
C^L_{iAX} &=& g_2 {m_{l_i} \over \sqrt{2} m_W cos\beta} (V)_{A2} R^\nu_{Xi}
\eea 
In the above $U$ and $V$ are the diagonalising matrices of chargino mass matrix $M_C$, 
$R^\nu$ is the diagonalising matrix of the sneutrino mass matrix, $M_{\tilde{\nu}}^2$. And
the indices $A$ and $X$ runs over the dimensions of the respective matrices 
($ A = 1,2$ for Charginos, $X = 1,2,3$ for sneutrinos), whereas $i$ as usual 
runs over the generations, $m_{l_i}$ is the mass of the $i$~th lepton and
rest of the parameters carry the standard definitions. \\
\noindent
(ii) Fermion-Sfermion-Neutralino :\\
In a similar manner, the fermion-sfermion-neutralino vertex is given by:
\be
\tilde{D}_{iAX} = D^R_{iAX} P_R + D^L_{iAX} P_L
\ee
where $D^L$ and $D^R$ have the following forms:
\bea
D^R_{iAX} &=& -{g_2 \over \sqrt{2}} \left\{ \left[ -N_{A2} - N_{A1} tan\theta_W \right] R^l_{Xi} + 
{m_{l_i} \over m_W cos\beta} N_{A3} R^l_{X,i+3}\right\} \\
D^L_{iAX} &=& -{g_2 \over \sqrt{2}} \left\{ {m_{l_i} \over m_W cos\beta} 
N_{A3} R^l_{Xi} + 2 N_{A1} tan\theta_W R^l_{X,i+3} \right\}
\eea
In the above $N$ is  diagonalising matrices of neutralino mass matrix $M_N$, $R^l$ is
the diagonalising matrix of the slepton mass matrix, $M_{\tilde{l}}^2$. And
the indices $A$ and $X$ runs over the dimensions of the respective matrices 
($A = 1,..,4$ for neutralinos, $X = 1,..,6$ for sleptons), whereas $i$ as usual 
runs over the generations.

\subsubsection{Think it Over:}
The LEP experiment at CERN searched for a light Higgs boson which has SM like couplings
through the process $e^+e^- ~\to~ZH$ and has a put a limit on the lightest Higgs
boson mass as $m_h ~\ger~ 114.2 $GeV. This limit applies to the light Higgs boson
of the MSSM (except in some range and in the presence of CP violation in the Higgs
sector). Take the formula of the 1-loop Higgs mass given by eq.(\ref{higgs1loop})
and simplify it by assuming the stop masses are of the similar order $~ M_S$ and 
the mixing between the stops is maximal. Find out what is the least value of the
$M_S$ which is consistent with the Higgs mass. Now compute the 1-loop corrections
to the minimisation conditions and check what is the amount of fine-tuning 
required to obtain the correct $M_Z$ mass.  Show that  a few percent fine tuning 
is already required to satisfy the LEP limit on the light Higgs mass.  The fine tuning
rapidly increases with increasing Higgs mass. This goes under the name \textit{Little
Hierarchy Problem}.

\section{`Standard' Models of Supersymmetry breaking}
So far we have included supersymmetry breaking within the MSSM through a set of 
explicit  supersymmetry breaking soft terms however, at a more fundamental we
would like to understand the origins of these soft terms as coming from a 
theory where supersymmetry is spontaneously broken. In a previous section, we
have mentioned that supersymmetry needs to be broken spontaneously in a 
hidden sector and then communicated to the visible sector through a messenger
sector.  In the below we will consider two main models for the messenger sector (a)
the gravitational interactions and (b) the gauge interactions. But before 
we proceed to list problems  with the general form  soft supersymmetry breaking terms
as discussed in the previous section. 
This is essential to understand what kind of constructions of supersymmetric
breaking models are likely to be realised in Nature and thus are consistent
with phenomenology.

The way we have parameterised supersymmetry breaking in the MSSM, using
a set of gauge invariant soft terms, at the first sight, seems to be
the most natural thing to do in the absence of a complete theory of
supersymmetry breaking. However, this approach is itselves laden with
problems as we realise once we start confronting this model with 
phenomenology. The two main problems can be listed as below: \\
\noindent
(i). Large number of parameters \\
\noindent
Compared to the SM, in MSSM, we have a set of more than 50 new particles; 
writing down all possible gauge invariant and supersymmetry breaking soft
terms, limits the number of possible terms to about 105. All these
terms are completely arbitrary, there is no theoretical input 
on their magnitudes, relative strengths, in short there is no
theoretical guiding  principle about these terms. Given that these are
large in number, they can significantly effect the phenomenology.
In fact, the MSSM in its softly broken form seems to have lost
predictive power except to say that there are some new particles within
a broad range in mass(energy) scale. The main culprit being the large
dimensional parameter space $\sim$ 105 dimensional space
which determines the couplings of the supersymmetric particles 
and their the masses. If there is a model of supersymmetry breaking
which can act as a guiding principle and reduce the number of 
free parameters of the MSSM, it would only make MSSM more predictive. 

\noindent
(ii). Large Flavour and CP violations.
\noindent
As mentioned previously, the soft mass terms $m_{ij}^2$ and the 
trilinear scalar couplings $A_{ijk}$ can violate flavour. This
gives us new flavour violating structures beyond the standard
CKM structure of the quark sector which can also be incorporated
in the MSSM. Furthermore, all these couplings can also be complex 
and thus could serve as new sources of CP violation in addition  to 
the CKM phase present in the Standard Model. Given that all these
terms arbitrary and could be of any magnitude close to weak scale,
these terms can contribute dominantly compared to the SM amplitudes
 to various  flavour violating processes at the weak scale, like  flavour 
violating decays like $b \to s + \gamma$ or flavour oscillations like 
$K^0 \leftrightarrow \bar{K}^0$ etc and even flavour violating decays
which do not have any Standard Model counterparts like 
$\mu \to e + \gamma$ etc. The CP violating phases can also contribute 
to electric dipole moments (EDM)s which are precisely measured at
experiments.

To analyse the phenomenological impact of these processes on these
terms, an useful and powerful tool is the so called
Mass Insertion (MI) approximation.
In this approximation, we use flavour diagonal gaugino
vertices and the flavour changing is encoded in non-diagonal sfermion
propagators. These propagators are then expanded assuming
that the flavour changing parts are much smaller than the flavour
diagonal ones. In this way we can isolate the relevant elements of the
sfermion mass matrix for a given flavour changing process and it is not
necessary to analyse the full $6\times 6$ sfermion mass matrix.
Using this method, the experimental limits lead to upper bounds on the
parameters (or combinations of) $\delta_{ij}^f \equiv 
\Delta^f_{ij}/m_{\tilde{f}}^2$, known as mass insertions;
where $\Delta^f_{ij}$ is the flavour-violating off-diagonal entry
appearing in the $f = (u,d,l)$ sfermion mass matrices and
$m_{\tilde{f}}^2$ is the average sfermion mass. In addition, the
mass-insertions are further sub-divided into LL/LR/RL/RR types,
labeled by the chirality of the corresponding SM fermions. 
The limits on various $\delta$'s coming from various flavour 
violating processes have been computed and tabulate in the literature 
and can be found for instance in Ref.\cite{ourlectures}. 

These limits show that the flavour violating terms should be typically
at least a couple of orders of magnitude suppressed compared to the flavour
conserving soft terms\footnote{The flavour problem could also be alleviated
by considering decoupling soft masses or alignment mechanisms.}. While this is true for the first two generations
of soft terms, the recent results from B-factories have started constraining
flavour violating terms involving the third generation too. In light of 
this stringent constraint, it is more plausible to think that the fundamental
supersymmetry breaking mechanism some how suppresses these flavour violating
entries. Similarly, this mechanism should also reduce the number of parameters
such that the MSSM could be easily be confronted with phenomenology and
make it more predictive.  We will consider two such models of supersymmetry
breaking below which will use two different kinds of messenger
sectors.

\subsection{Minimal Supergravity}
In the minimal supergravity framework, gravitational interactions 
play the role of messenger sector. Supersymmetry is broken spontaneously
in the hidden sector. This information is communicated to the
MSSM sector through gravitational sector leading to the soft terms. 
Since gravitational interactions play an important role only at 
very high energies, $M_{p} \sim O(10^{19})$ GeV, the breaking 
information is passed on
to the visible sector only at those scales. 
The strength of the soft terms is characterised roughly by, 
$m_{\tilde{f}}^2 ~\approx~  M_S^2 / M_{planck}$, where $M_S$ is the scale of supersymmetry
breaking. These masses can be comparable to weak scale for $M_S \sim 10^{10}$ GeV.  
This $M_{S}^2$ can correspond to the F-term vev of the Hidden sector.
The above mechanism of supersymmetry breaking is called 
supergravity (SUGRA) mediated supersymmetry breaking. 

A particular class of supergravity mediated supersymmetry breaking
models are those which go under the name of "minimal" supergravity. 
This model has special features that it reduces to total number
of free parameters determining the entire soft spectrum to five.
Furthermore, it also removes the dangerous flavour violating
soft terms in the MSSM. The classic features of this model are
the following boundary conditions to the soft terms at the high
scale $\sim ~M_{Planck}$ :
\begin{itemize}
\item All the gaugino mass terms are equal at the high scale.
	$$M_1 = M_2 = M_3 = M_{1/2}$$
\item All the scalar mass terms at the high scale are equal.
	$$m_{\phi_{ij}}^2 = m_0^2 \delta_{ij}$$
\item All the trilinear scalar interactions are equal at the high
	scale. 
	$$A_{ijk} = A h_{ijk}$$
\item All bilinear scalar interactions are equal at the high scale. 
	$$B_{ij} = B$$ 	
\end{itemize}
Using these boundary conditions, one evolves the soft terms to 
the weak scale using renormalisation group equations. 
It is possible to construct supergravity models which can give rise
to such kind of strong universality in soft terms close to Planck scale. 
This would require the Kahler potential of the theory to be of the 
canonical form. As mentioned earlier, the advantage of this model 
is that it drastically 
reduces the number of parameters of the theory to about five, 
$m_0, M$~(or equivalently $M_2$), ratio of the {\it vevs} of the two 
Higgs, tan$\beta$, $A$, $B$. Thus, these models
are also known as `Constrained' MSSM in literature. 
The supersymmetric mass spectrum of these models has been extensively 
studied in literature. The Lightest Supersymmetric
Particle (LSP) is mostly a neutralino in this case. 

\subsection{Gauge Mediated Supersymmetry breaking}

 In a more generic
case, the Kahler potential need not have the required canonical form.
In particular, most low energy effective supergravities from string theories
do not posses such a Kahler potential. In such a case, 
large FCNC's and again large number of parameters are expected from 
supergravity theories. 
An alternative mechanism has been proposed which tries to avoid
these problems in a natural way. The key idea is to use gauge interactions
instead of gravity to mediate the supersymmetry breaking from the 
hidden (also called secluded sector sometimes) to the visible MSSM sector.
In this case supersymmetry breaking can be communicated 
at much lower energies $\sim 100$ TeV. 

A typical model would contain a susy breaking sector called `messenger 
sector' which contains a set of superfields transforming under a gauge
group which `contains' $G_{SM}$. Supersymmetry is broken spontaneously 
in this sector and this breaking information is passed on to the 
ordinary sector through gauge bosons and their fermionic partners in loops. 
The end-effect of this mechanism also is to add the soft terms in to the
lagrangian. But now these soft terms are flavour diagonal as they are
generated by gauge interactions. The soft terms at the messenger scale 
also have simple expressions in terms of the susy breaking parameters. 
In addition, in minimal models of gauge mediated supersymmetry breaking, 
only  one parameter can essentially determine the entire soft spectrum. 

In a similar manner as in the above, the low energy susy spectrum is 
determined by the RG scaling of the soft parameters. But now the high 
scale is around 100 TeV instead of $M_{GUT}$ as in the previous case. 
The mass spectrum of these models has been studied in many papers.
The lightest supersymmetric particle in this case is
mostly the gravitino in contrast to the mSUGRA case. 

\subsubsection{Think it Over}
\begin{itemize}
\item In both gravity mediated as well as gauge mediated supersymmetry
breaking models, we have seen that RG running effects have to 
included to study the soft terms at the weak scale. Typically,
the soft masses which appear at those scales are positive at the 
high scale. But radiative corrections can significantly modify 
the low scale values of these parameters; in particular, 
making one of the Higgs mass to be negative at the weak scale leading
to spontaneous breaking of electroweak symmetry. This 
mechanism is called radiative electroweak symmetry breaking. Consider
two hypothetical situations when (a) the top mass is twice its 
present value $m_t~ = 2~ m_t $ (b) the top mass is 1/10 th its
present value $m_t~ = ~ m_t/10 $. In which case there would be 
more efficient Electroweak symmetry breaking ?
\item  The recent limits from LHC already put severe constraints 
on the lightest squarks and gluino masses. They push their masses 
to be greater  than 800 GeV - 1 TeV.  In fact, this has severe constraints
on mSUGRA model. For latest limits have a look at \cite{cerntwiki}. 
\end{itemize}

\section{Remarks}
The present set of lectures are only a set of elementary introduction to 
the MSSM. More detailed accounts can be found in various references 
which we have listed at various places in the text. In preparing 
for these set of lectures, I have greatly benefitted from various
review articles and text books. I have already listed some of them
at various places in the text.
Martin's review \cite{martin} is perhaps the most comprehensive and
popular references. It is also constantly updated.  Some other excellent
reviews are  \cite{peskin1} and  \cite{bagger}.  A concise
introduction can also be found in \cite{csaki}. For more formal aspects
of supersymmetry including a good introduction to supergravity please
have a look at \cite{vanproeyen} and \cite{west}. For Grand Unified
theories and supersymmetry, please have a look at \cite{mohapatra},
\cite{yanagida}, \cite{ramond} and \cite{rosstextbook}. For a comprehensive
introduction to supersymmetric dark matter, please see \cite{kamionkowski}.
Finally, I would also recommend the original papers of anomaly mediated
supersymmetry breaking \cite{amsb}. Happy Susying.


\textbf{Acknowledgements} 
We would like to thank Urjit Yajnik for detailed discussions and consultations
through out the teaching period. We would also like to thank Ranjan Laha and
Manimala Mitra for going through these lecture notes several times and pointing
out various typographical errors. The author is supported by DST Ramanujan 
fellowship and DST project "Complementarity between direct and indirect searches
of supersymmetry".

\end{document}